\documentclass{fmj2010}
\usepackage{graphicx}
\begin{document}
   \title{Sub-parsec scale imaging of \mbox{Centaurus A}}

   \author{Cornelia M\"uller\inst{1},
		M. Kadler\inst{1,2,3},
		R. Ojha\inst{4,5},
                    M. B\"ock\inst{1},
	C. M. Fromm\inst{7},
	 E. Ros\inst{6,7},
	R. E. Rothschild\inst{8},
	\mbox{J. Wilms}\inst{1}
\and the TANAMI team
          }

   \institute{Dr. Remeis-Sternwarte \& ECAP, Sternwartstrasse 7, 96049 Bamberg, Germany
         \and
		CRESST/NASA Goddard Space Flight Center, Greenbelt, MD 20771, USA
          \and
		USRA, 10211 Wincopin Circle, Suite 500 Columbia, MD 21044, USA
	\and
		United States Naval Observatory, 3450 Massachusetts Ave., NW, Washington DC 20392, USA
	\and
		NVI,Inc., 7257D Hanover Parkway, Greenbelt, MD 20770, USA
	\and	Departament d'Astronomia i Astrofisica, Universitat de Val\`encia, E-46100 Burjassot, Spain
	\and	Max-Planck-Institut f\"ur Radioastronomie, Auf dem H\"ugel 69, 53121 Bonn, Germany
	\and	Center for Astrophysics and Space Sciences, University of California, San Diego, 9500 Gilman Drive, La Jolla, CA 92093-0424, USA
             }

   \abstract{

At a distance of about $3.8$\,Mpc, the radio galaxy Centaurus~A is the closest active galaxy. Therefore it is a key target for studying the innermost regions of active galactic nuclei (AGN). 
VLBI observations conducted within the framework of the TANAMI program enable us to study the central region of the Cen A jet with some of the highest linear resolutions ever achieved in an AGN. This region is the likely origin of the $\gamma$-ray emission recently detected by the \textsl{Fermi} Large Area Telescope (LAT).
TANAMI monitors a sample of radio and $\gamma$-ray selected extragalactic jets south of $-30^{\circ}$ declination at 8.4\,GHz and 22.3\,GHz with the Australian Long Baseline Array (LBA) and the transoceanic antennas Hartebeesthoek in South Africa, the 6\,m Transportable Integrated Geodetic Observatory (TIGO) in Chile and the 9\,m German Antarctic Receiving Station (GARS) in O'Higgins, Antarctica. The highest angular resolution achieved at 8.4\,GHz in the case of Cen~A is 0.59\,mas $\times$ 0.978\,mas (natural weighting) corresponding to a linear scale of less than 18 milliparsec. 

We show images of the first three TANAMI $8.4$\,GHz observation epochs of the sub-parsec scale jet-counterjet system of Cen~A. With a simultaneous 22.3\,GHz observation in 2008 November, we present a high resolution spectral index map of the inner few milliarcseconds of the jet probing the putative emission region of $\gamma$-ray-photons.
   }

 \authorrunning{C. M\"uller et al.}
  \titlerunning{Sub-parsec scale imaging of \mbox{Centaurus A}}
   \maketitle
%

\section{Introduction}
Centaurus A (PKS 1322-427) is the closest active radio galaxy at a distance of $3.8 \pm 0.1$\,Mpc (\cite{Harris2009}), where an angular resolution of one milliarcsecond (mas) corresponds to $\sim 0.018$\,pc. The optical counterpart of Cen~A is a giant elliptical galaxy (NGC 5128) which hosts a supermassive black hole with a mass \mbox{$M=5.5\pm3.0\times 10^7 \mathrm{M}_{\sun}$} (\cite{Israel1998, Neumayer2010}).  

Due to its proximity, Cen~A is an exceptionally good laboratory for studying the innermost regions of active galactic nuclei (AGN).
Cen~A can be seen over the whole range of the electromagnetic spectrum up to highest energies. Recently, the $\gamma$-ray detection by \textit{CGRO}/EGRET was confirmed by \textsl{Fermi}/LAT (\cite{Hartman1999, Abdo2010b}). In the TeV range, Cen~A was detected by H.E.S.S. (\cite{Aharonian2009}).

The radio source Cen~A is usually classified as a Fanaroff-Riley type I (FR~I) radio galaxy (\cite{Fanaroff1974}). The spectrum of the core ($\leq 4$\,mas) is inverted, indicating synchrotron or even free-free self absorption (Tingay et al.~\cite{Tingay1998b}). On subparsec scales, the radio jet-counterjet system was clearly resolved with the VLBI Space Observatory Program (VSOP) by Horiuchi et al. (\cite{Horiuchi2006}) at 5\,GHz.
As part of the TANAMI program\footnote[1]{Tracking Active Galactic Nuclei with Austral Milliarcsecond Interferometry\\ http://pulsar.sternwarte.uni-erlangen.de/tanami}, we produce images of Cen A at comparable resolution with only ground based telescopes.
Here we give an overview of all TANAMI observations of Cen~A prior to November 2008 and first results of the ongoing analysis.

\section{Observations and Data reduction}
\begin{figure*} 
   \centering
  \includegraphics[width=\textwidth]{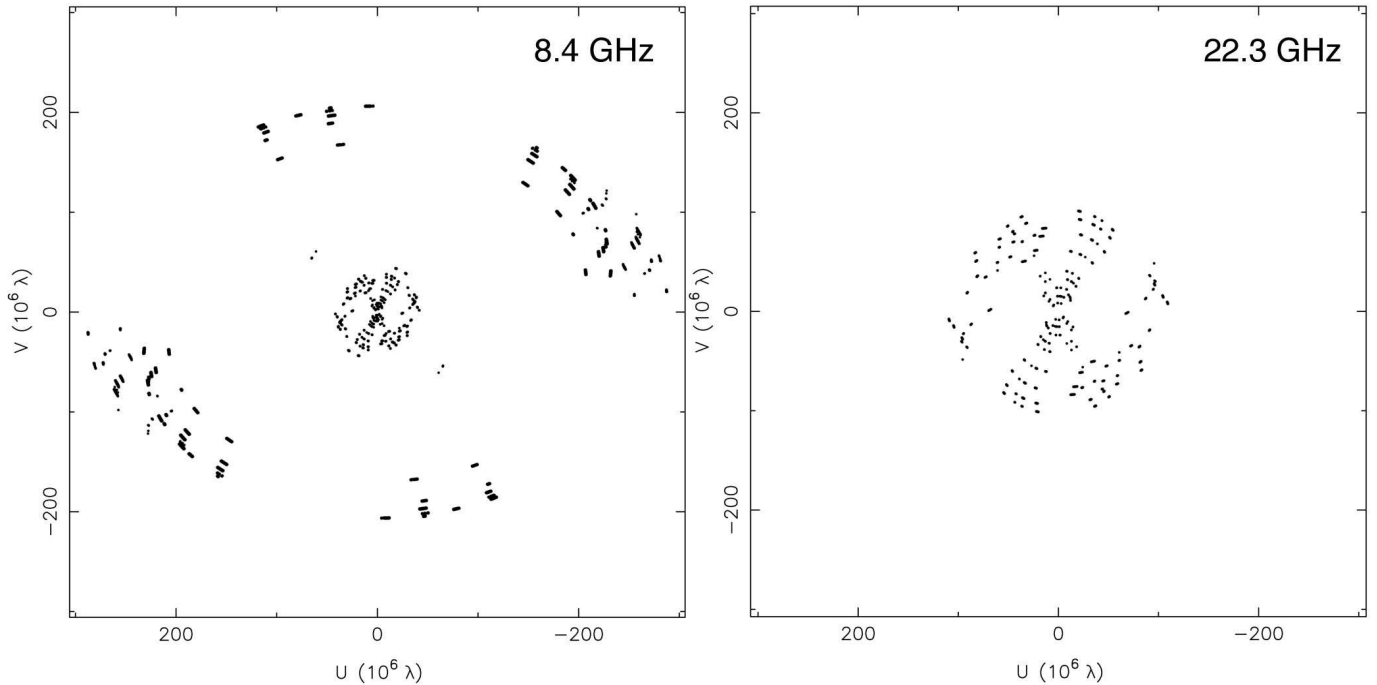}
  \caption{($u$-$v$)-coverage at 8.4\,GHz (left) and 22.3\,GHz (right) for Centaurus A}
              \label{fig:Fig1}
    \end{figure*}

\begin{table*}
\caption{Image parameters and observation characteristics (natural weighting)} 
\label{table:1} 
\centering 
\begin{tabular}{c c c c c c c c} 
\hline\hline 
Frequency & Epoch & RMS $^\ast$& $S_{\mathrm{peak}}$ & $S_{\mathrm{total}}$ & $\theta_{\mathrm{maj}}$ &  $\theta_{\mathrm{min}}$ & P.A. \\
 $[\mathrm{GHz}]$	& yyyy-mm-dd& [$\mathrm{mJy\,beam^{-1}}$] & [$\mathrm{Jy\,beam^{-1}}$] & [$\mathrm{Jy}$] & [$\mathrm{mas}$] & [$\mathrm{mas}$] & [$^\circ$]\\
\hline 
8.4 & 2007-11-10 &0.37	$\pm 0.04$ &0.60	& 2.6	$\pm 0.1$ &1.64	&0.41	&	7.9 \\ 
8.4 & 2008-06-09 &0.58 $\pm 0.05$ &1.06	&  3.1 $\pm 0.1$ &   2.86 &1.18	&$-12.7$	 \\ 
8.4 & 2008-11-27 & 0.39	$\pm 0.02$ &0.74	&3.9	$\pm 0.1$ &0.98	&0.59 &31.4	\\ 
22.3 & 2008-11-29 &0.47 $\pm 0.03$	&1.77	& 3.4	$\pm 0.1$ &2.01	&1.27 & 87.9	 \\
\hline 
\end{tabular}\\
\footnotesize{$^\ast$ RMS values are determined in a region of the final map without significant source flux.}
\end{table*}
The Very Long Baseline Interferometry (VLBI) observations of Cen~A presented here were made in the framework of the TANAMI program (Ojha et al. \cite{Ojha2010}). The participating telescopes in this southern-hemisphere VLBI project are the Australian Long Baseline Array (LBA), with antennas in Narrabri (5$\times$22\,m), Ceduna (30\,m), Hobart (26\,m), Mopra (22\,m), Parkes (64\,m), the 70\,m and 34\,m telescopes of NASA's Deep Space Network (DSN) located at Tidbinbilla, the $26\,$m South-African Hartebeesthoeck antenna, the 9\,m German Antarctic Receiving Station (GARS) in O'Higgins, Antarctica, and the 6\,m Transportable Integrated Geodetic Observatory (TIGO) in Chile. The TANAMI source list consists currently of 75 extragalactic jets. Observations are conducted at 8.4\,GHz and 22.3\,GHz (for more details see Ojha et al., these proceedings). The $\gamma$-ray properties of the sample including Cen~A are presented by B\"ock et al. in these proceedings.

Figure~\ref{fig:Fig1} shows the ($u$-$v$)-coverage for Cen~A at 8.4\,GHz (left) and 22.3\,GHz (right) for the November 2008 TANAMI observation. The radial ($u$-$v$)-coverage at 170--290\,M$\lambda$ is provided by baselines involving the TIGO and O'Higgins antennas. This results in an angular resolution of \mbox{0.59\,mas $\times$ 0.978\,mas} (natural weighting) at 8.4\,GHz.

Since the transoceanic antennas O'Higgins and TIGO do not support observations at 22.3\,GHz, at this frequency the resolution is lower (\mbox{2.01\,mas $\times$ 1.27\,mas}). 

Data calibration and hybrid imaging were performed by using standard techniques as described by Ojha et al. (\cite{Ojha2010}). 

\section{Results}
The image parameters and observation characteristics of the three 8.4\,GHz and the one 22.3\,GHz TANAMI observations of Cen~A are listed in Table~\ref{table:1}. 
\begin{figure}
\centering
\includegraphics[width=0.88\columnwidth]{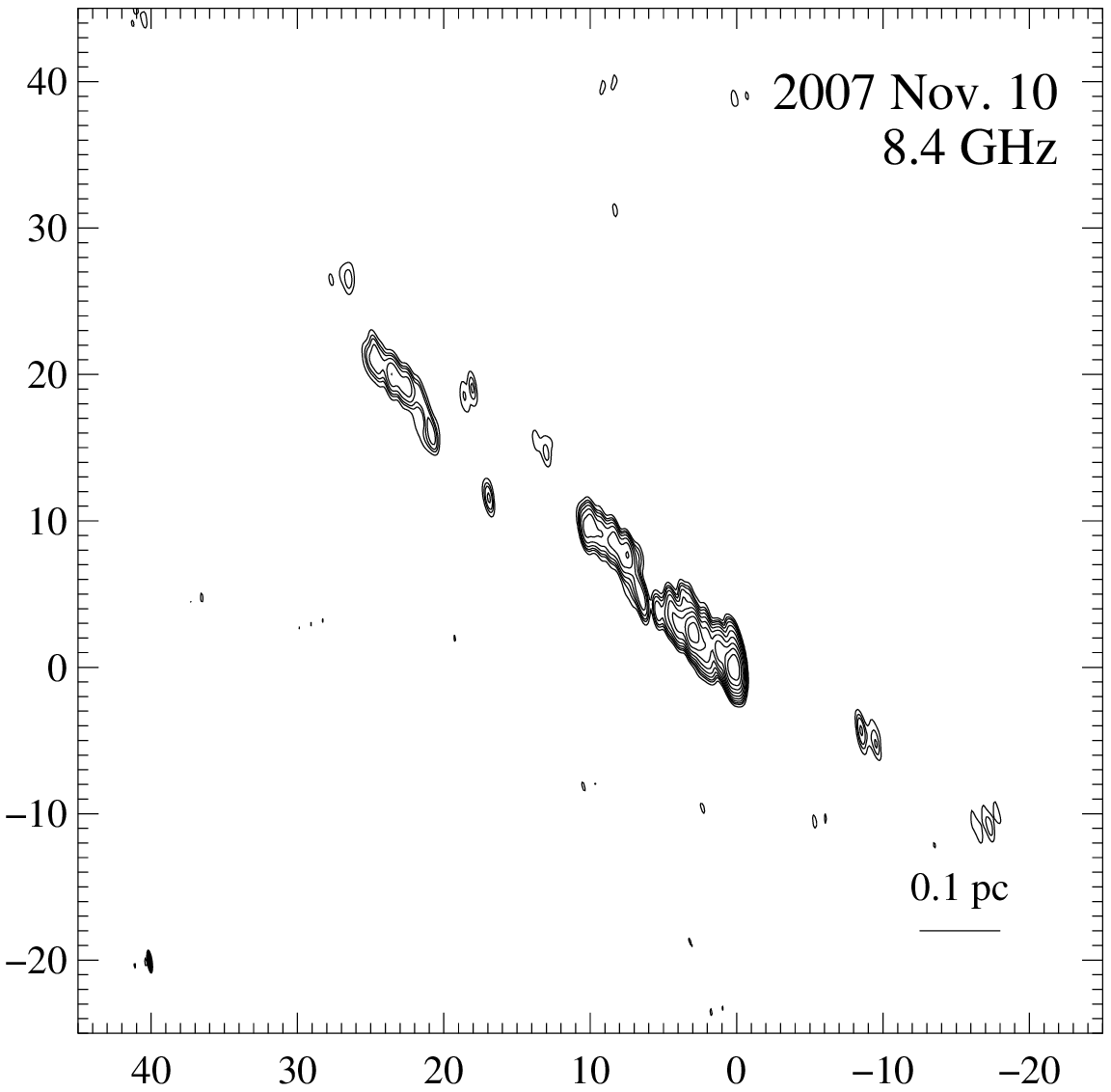}
\includegraphics[width=0.88\columnwidth]{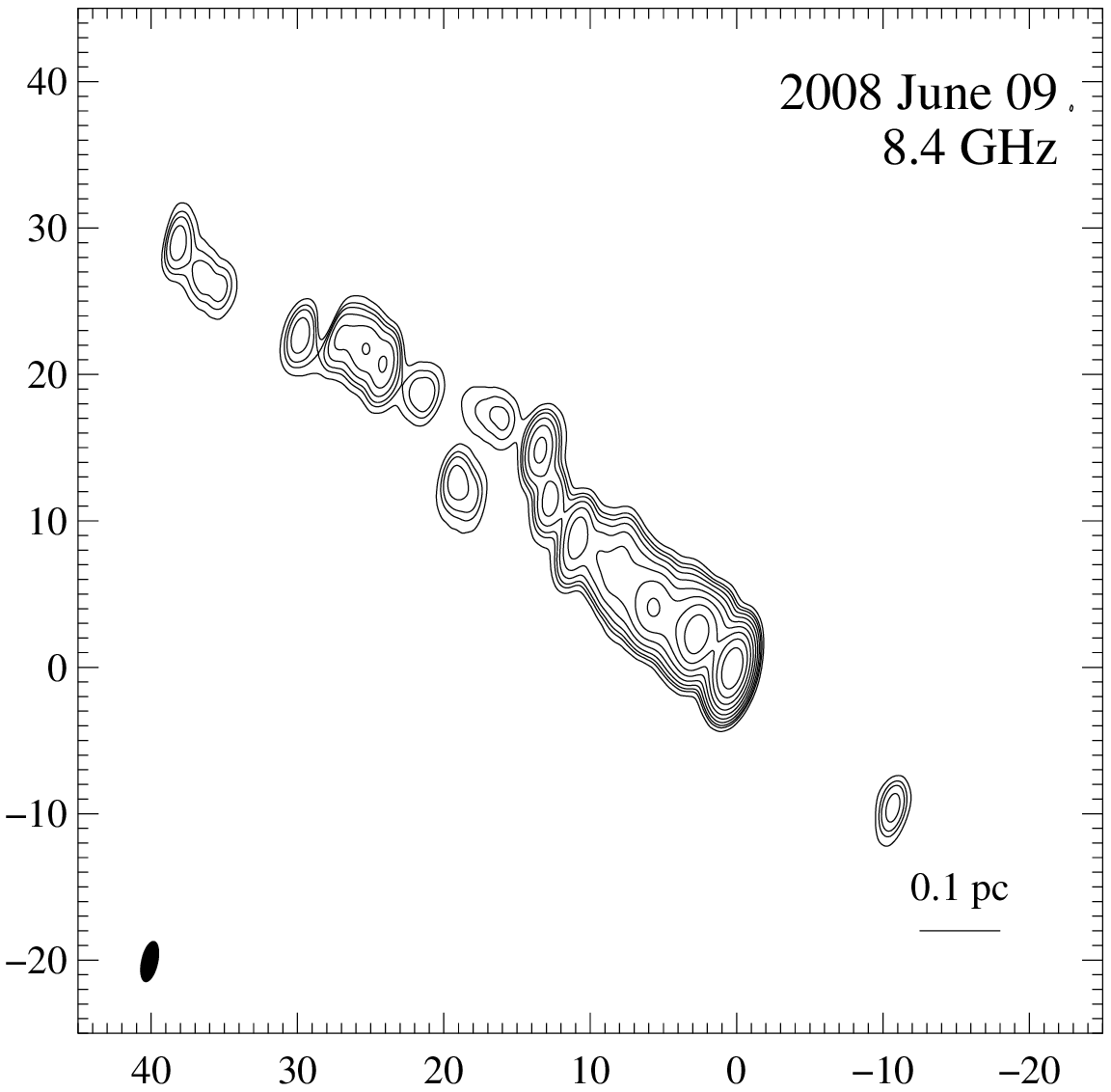}
\includegraphics[width=0.88\columnwidth]{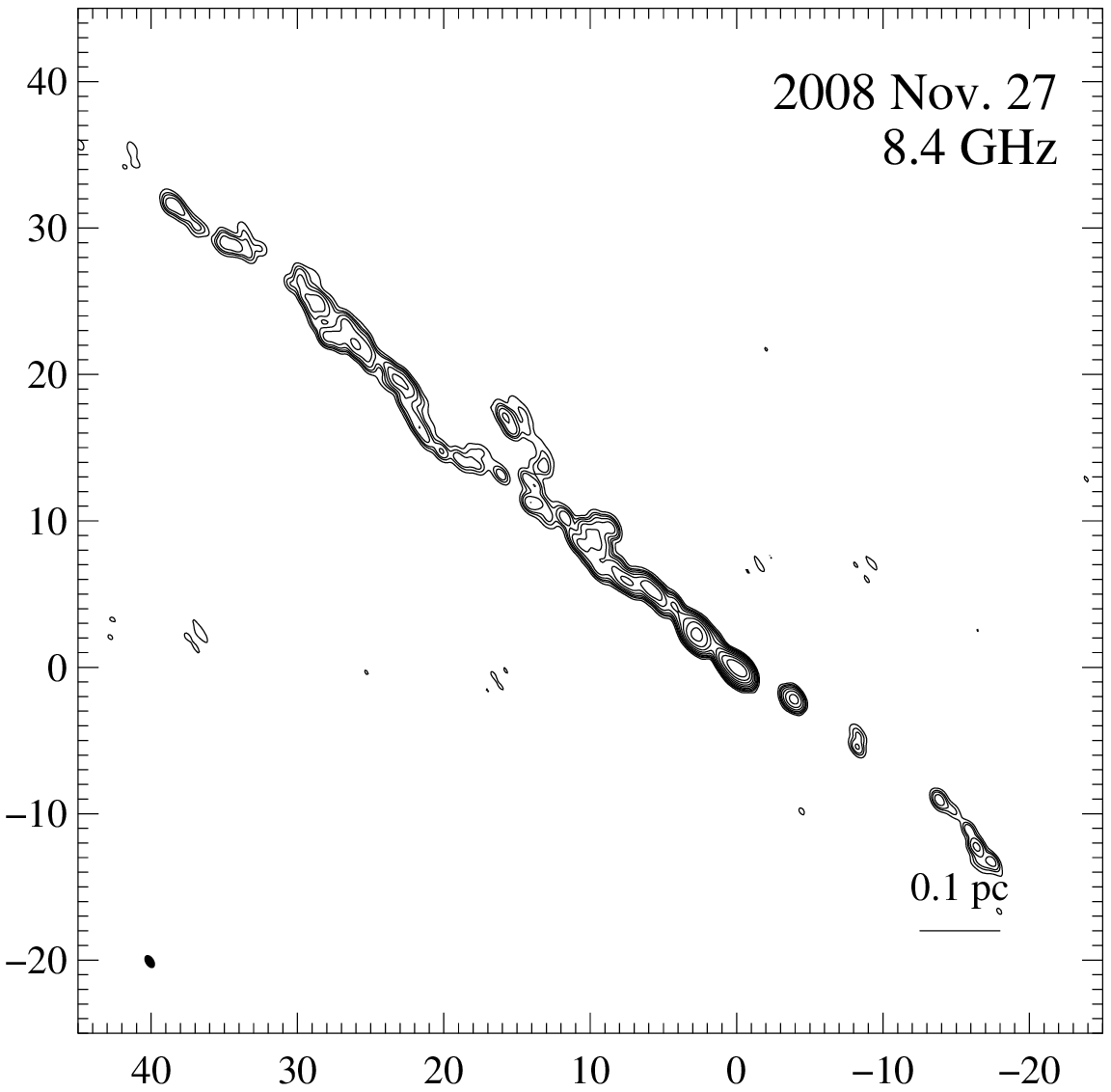}
\caption{8.4\,GHz images of Cen~A of November 2007 (top), June 2008 (middle) and of November 2008 (bottom). The lowest contours are at $3\sigma$.}
         \label{fig:Fig2}
\end{figure}
Imaging was performed with the program \textsc{difmap} (\cite{Shepherd1997}), using the \textsc{clean} algorithm and making use of phase and amplitude self-calibration. 

\begin{figure}
   \centering 
\includegraphics[width=\columnwidth]{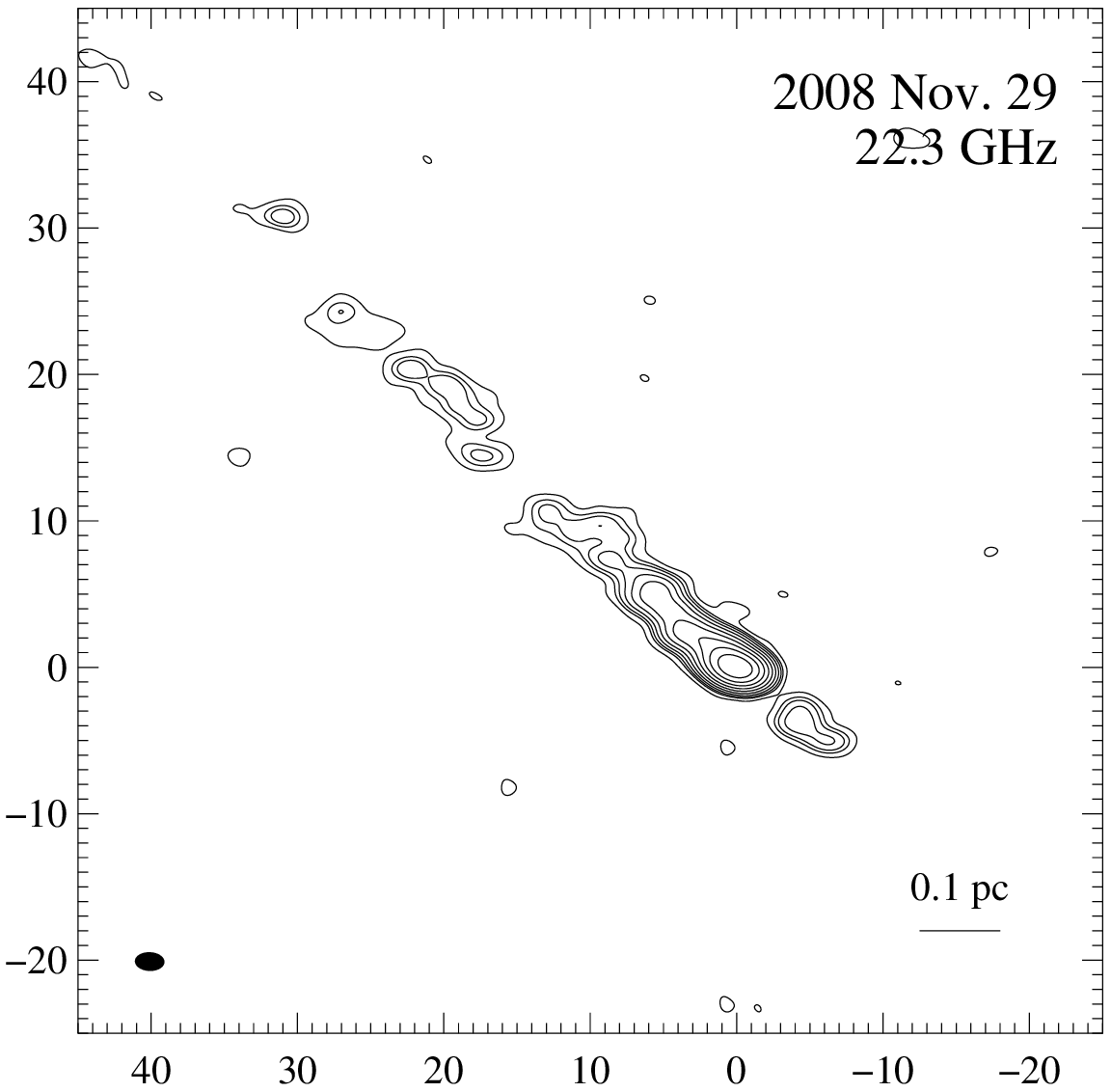}
\caption{22.3\,GHz image of November 2008}
\label{fig:Fig3}
\hspace{3mm}
  \includegraphics[width=\columnwidth]{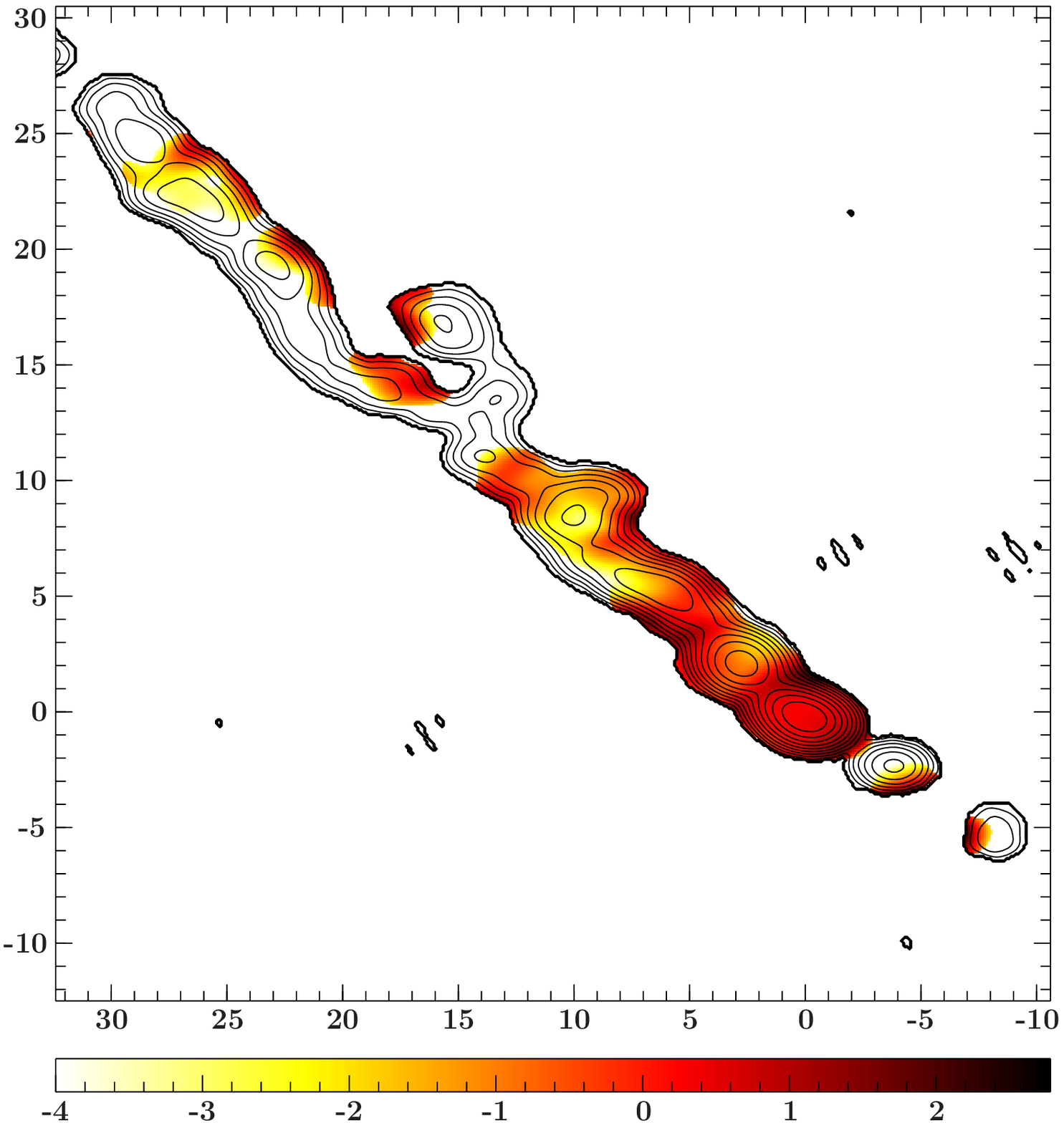}
  \caption{Spectral index map as calculated for $S_{\mathrm{8.4GHz}}\geq 3\sigma_{\mathrm{8.4GHz}}$ and $S_{\mathrm{22.3\,GHz}}\geq 2\sigma_{\mathrm{22.3\,GHz}}$. The overlaying contours show the flux density distribution at 8.4\,GHz folded with a common beam of $1.61 \times 1.02$\,mas (P.A. $=88^\circ$). The spectral index is defined as \mbox{$F_\nu \sim \nu^{+\alpha}$}.}
              \label{fig:Fig4}
\end{figure}

TANAMI monitored Cen~A until November 2008 three times at 8.4\,GHz and one time simultaneously at 22.3\,GHz. Figures~\ref{fig:Fig2} and \ref{fig:Fig3} show the resulting naturally weighted 8.4\,GHz and 22.3\,GHz images. With only ground based telescopes, we achieve highest angular resolution observations of Cen~A, which can be compared with earlier space-VLBI observations (Horiuchi et al.~\cite{Horiuchi2006}). The smallest resolved structures are on the scale of 12 light-days.

We resolve the core region of Cen~A into several jet components. At both frequencies and at all epochs, a well collimated jet at a mean position angle (P.A.) of $\sim 50^\circ$ and a fainter counterjet (P.A.$\sim -130^\circ$) with an emission gap in between is seen. The significant features within the sub-parsec scale jet observed in the November 2007 image at 8.4\,GHz of TANAMI observations  (\cite{Ojha2010}) are in good agreement with those of the following epochs. This result can be used to set constraints on the position of the core. 

In the highest resolution image of 2008 November, a widening of the jet at about 25\,mas \mbox{($\approx 0.4$\,pc)} downstream and a subsequent recollimation are observed. These features appear also in the November 2007 and June 2008 images as well as in the 22.3\,GHz map at lower resolution (see Figs.~\ref{fig:Fig2} \&~\ref{fig:Fig3}).

The November 2008 image reveals a small counterjet displacement from the jet line, which can also be seen in the image of Horiuchi et al. (\cite{Horiuchi2006}).

Both, the peak and the total flux densities at the 8.4\,GHz epochs show only moderate variability with a mean of $S_{\mathrm{peak}} \sim 0.71\,\mathrm{Jy\,beam^{-1}}$ and $S_{\mathrm{total}} \sim 3.5\,\mathrm{Jy}$. At 22.3\,GHz, the flux density is higher indicating an inverted core spectrum (see below).

By analyzing $\sim$\,8 years of observations at multiple frequencies Tingay et al.~(\cite{Tingay1998b}) measured a jet speed of 0.1\,c. Our three 8.4\,GHz jet images (separated by 0.5 years each) can be fitted with a self-consistent model of Gaussian components within the inner 25\,mas. At least one more observation epoch is required to measure robust component velocities. 

Despite a larger synthesized beam at the higher frequency, the jet structures at 8.4\,GHz and 22.3\,GHz in the simultaneous measured November 2008 images match well within the inner 30\,mas ($\approx 0.5$\,pc) of the jet. We modeled the core at both frequencies with three Gaussian components. The comparison of the optically thin components reveals a shift of the 22.3\,GHz core with respect to the 8.4\,GHz core in the direction of the central black hole of $\Delta\alpha \approx 0.2$\,mas and $\Delta\delta\approx0.15$\,mas. Taking this alignment correction into account, we obtained the spectral index distribution along the jet shown in Fig.~\ref{fig:Fig4}. Both images were restored with a common beam ($80\%$ of synthesized beam at 22.3\,GHz; $1.61 \times 1.016$\,mas, P.A. 88$^\circ$). The overlaid contours correspond to the 8.4\,GHz image folded with this common beam. The core region has an inverted spectrum which changes from flat to steep downstream. The highest spectral indices with values $\alpha \leq 1$ are found in the core, indicating synchrotron self-absorption. 

Recently, the  \textsl{Fermi} Large Area Telescope detected $\gamma$-ray emission from the Cen~A lobes (\cite{2010Sci...328..725F}), as well as from its core (Abdo et al. 2010b, submitted). Flat spectrum regions in the sub-parsec scale radio jet are possible production regions of high energetic photons (\cite{Marscher2010}). We identify the inner few milliarcseconds ($\sim 0.2$\,pc) at 8.4\,GHz of the Cen~A radio jet as possible sources of $\gamma$-ray emission with the strongest inverted-spectral emission coming from the jet core on scales of $\leq0.1$\,pc.

\section{Conclusions}
We presented the first TANAMI observations of Cen~A at 8.4\,GHz and 22.3\,GHz including the highest resolved image of Cen~A ever made with ground based telescopes. With a simultaneous observation in November 2008 we were able to produce a spectral index map of the milliparsec-scale jet of Cen~A identifying the putative $\gamma$-ray emission regions.

Further analysis of the following TANAMI observations of Cen~A will try to test the previously determined jet speeds.

\begin{acknowledgements}
We thank the rest of the TANAMI team for their collaboration. 

The Long Baseline Array is part of the Australia Telescope which is funded by the Commonwealth of Australia for operation as a National Facility managed by CSIRO. This work made use of the Swinburne University of Technology software correlator, developed as part of the Australian Major National Research Facilities Programme and operated under licence. 

This research has made use of NASA’s Astrophysics Data System and the NASA/IPAC Extragalactic Database (NED, operated by the Jet Propulsion Laboratory, California Institute of Technology, under contract with the National Aeronautics and Space Administration).

We acknowledge partial support by the Bundesministerium f\"ur Wirtschaft und Technologie through Deutsches Zentrum f\"ur Luft- und Raumfahrt grant 50 OR 0808. 
\end{acknowledgements}
%

\begin{thebibliography}{}
\bibitem[Fermi-LAT Collaboration 2010]{2010Sci...328..725F} Fermi-LAT 
Collaboration 2010, Science, 328, 725 

\bibitem[Abdo et al. 2010]{Abdo2010a} Abdo, A.~A., et al.\ 2010, submitted to ApJ,
arXiv:1005.2626 

\bibitem[Abdo et al. 2010]{Abdo2010b} Abdo, A.~A., et al.\ 2010, 
\apjs, 188, 405

\bibitem[Aharonian et al. 2009]{Aharonian2009} Aharonian, F., et 
al.\ 2009, \apjl, 695, L40 

\bibitem[Fanaroff \& Riley 1974]{Fanaroff1974} Fanaroff, B.~L., \& Riley, J.~M.\ 1974, \mnras, 167, 31P

\bibitem[Harris et al. 2009]{Harris2009} Harris, G.~L.~H., 
Rejkuba, M., \& Harris, W.~E.\ 2009, arXiv:0911.3180 

\bibitem[Hartman et al. 1999]{Hartman1999} Hartman, R.~C., et al.\ 
1999, \apjs, 123, 79

\bibitem[2006]{Horiuchi2006} Horiuchi, S., Meier, 
D.~L., Preston, R.~A., \& Tingay, S.~J.\ 2006, \pasj, 58, 211 

\bibitem[Israel 1998]{Israel1998} Israel, F.~P.\ 1998, \aapr, 8, 237 

\bibitem[Marscher et al. 2010]{Marscher2010} Marscher, A.~P., Belloni, T.\ 2010, 
The Jet paradigm - From Microquasars to Quasars, Lecture Notes in Physics, 794, 173

\bibitem[Neumayer et al. 2010]{Neumayer2010} Neumayer, N., 
Cappellari, M., van der Werf, et al.\ 2010, The Messenger, 139, 36 

\bibitem[2010]{Ojha2010} Ojha, R., Kadler, M., B\"ock, M. et al.\ 2010, in press 
(arXiv:1005.4432)


\bibitem[Shepherd 1997]{Shepherd1997} Shepherd, M.~C.\ 1997, 
Astronomical Data Analysis Software and Systems VI, 125, 77 

\bibitem[1998]{Tingay1998b} Tingay, S.~J., Jauncey, D.~L., Reynolds, J.~E., et al.\ 
1998, \aj, 115, 960 


\end{thebibliography}
\end{document}